\documentclass[12pt]{article}

\def\title{Non decoupling ghosts in the light cone gauge.}

\long\def\abstract{The gist of using the light cone gauge lies
in the well known property of ghosts decoupling. 
But from the BRST point of view this is a stringency since for the
construction of a nilpotent operator (from a Lie algebra) the
presence of ghosts are mandatory. 
We will show that this is a foible which has its origins in
the very fact of using just one light cone vector ($n_\mu$) instead of working 
with both light cone vectors ($n_\mu$ and $m_\mu$) to fulfill the light 
cone base vectors. This will break out ghost decoupling from theory 
but allowing now a consistent BRST theory for the light cone gauge.
\\ \\
{\bf PACS: 02.90.+p; 12.38.Bx}}

\def\ket#1{|#1\!\!>}

\input{axodraw.sty}
\input{epsf}

\begin{document}

\vskip 15mm
\centerline{\huge\bf Non decoupling ghosts}
\centerline{\huge\bf in the light cone gauge.}
\vskip 8mm
\begin{center} \vskip 10mm
Ricardo Bent\'{\i}n{\begingroup\def\thefootnote{$\aleph$}
                     \footnote{e-mail: rbentin@uesc.br}
                     \addtocounter{footnote}{-1} \endgroup}
\\[3mm] Se\c c\~ao de Matem\'atica.\\
        Universidade Estadual de Santa Cruz.\\
        Rodovia Ilh\'eus-Itabuna Km. 18,  Ilh\'eus, BA, Brazil
\vskip 3.0cm              {\bf ABSTRACT } \end{center}    \abstract

\thispagestyle{empty} \newpage
\pagestyle{plain} 

\newpage
\setcounter{page}{1}

\baselineskip 0.7cm
BRST symmetry has a special delighting importance in theoretical physics.
The light cone gauge has also a very strong relevance in theoretical physics.\\
But meanwhile the former makes mandatory the use of ghosts \cite{siegel}, 
the later decoples them from theory \cite{josif} strangling the possibility
of building up a consistent BRST operator.\\ 
There is an interesting example concerning the use of both ideas in the context 
of string theory quantization \cite{big,gsw}. The (super)string quantization for
excellence is done making use the non covariant light cone gauge, this breaks
the covariance of the theory but it is a consistent and also the quickest way of 
do that. Hence there is a well succeeded attempt of obtain 
a covariant quantization of superstring 
using the ideia of mixing pure spinors with BRST operator \cite{nb},
in order to obtain a covariant quantization of superstring theory 
building up a nilpotent operator.\\
The history of light cone gauge theories is shoved with really
true degrees of freedom as was pointed out by Dirac \cite{dirac}.
The decoupling of ghosts in light cone gauge theories should
be felt as a great bonus; but there are some big prices
to pay for: a complicated structure of gauge propagator, 
the so called spurious poles appear in theory and the absence of 
consistent BRST operators (since ghosts are mandatory in 
constructing them \cite{bass}); just to cite some of them.\\
In order to show our ideas, we will use a non-abelian and 
also a non-covariant Yang-Mills theory in four dimensions since we want to
fix the gauge freedom on the light-cone.
This work is presented in four sections. The first one has a very briefly
review of BRST symmetry. Section two shows the {\it one} base light cone vector  
reaches the ghost decoupling. Section three will explain 
how the {\it completely two} base light cone vectors do not reach 
the ghost decoupling. Conclusions are written in last section.
\section{\large BRST operator gist.}
The BRST symmetry origins remotes to quantum field theory where it 
was discovered. But this symmetry is more than a quantum 
field property. It is associated to gauge degrees of freedom of 
constrained systems. 
For a Lie algebra defined in terms of is generators $G_a$ 
and structures constant $f_{ab}^c$ by
$$
  [G_a,G_b]=if_{ab}^cG_c,
$$ 
an operator, called the BRST operator is defined as
\begin{equation}
  \label{BRST}
  Q=c^aG_a-\frac{i}{2}f_{ab}^cc^ac^bb_c,
\end{equation}
where the $c^a$ and $b_a$ form a canonical conjugate pair of
anti commuting variables that are known as ghosts since its statistical
property and satisfy
$$
  \{b_a,c^b\}=\delta_a^b.
$$
The BRST operator is nilpotent, that is
$$
   \{Q,Q\}\equiv 2Q^2=0,
$$
which is easy to demonstrate making use of Jacobi identity. 
Also we could define the ghost number operator $J$ as
$$
  J=c^ab_a,
$$
that implies following relation $[J,Q]=Q$.
The very importance in defining the BRST operator is related to the 
fact that its cohomology give us the true physical states, 
say $\ket{\phi}$. If $\ket{\phi'}$ represents the same physical state, then it
is related $\ket{\phi}$ throughout
$$
  \ket{\phi'}=\ket{\phi}+Q\ket{\varphi},
$$
where $\ket{\varphi}$ is an arbitrary vetor. In other words, 
they belongs to the same cohomology since $Q\ket{\phi}=Q\ket{\phi'}$.
This fact can be used to classified the space into
equivalent classes. This condensate introduction on BRST operator
is that we need to have in mind for next sections.
\section{\large One light cone base vector.}
The light cone gauge Yang-Mills theory \cite{leibb} is practically 
defined in terms of its gauge fixing Lagrangian part
$$
{\cal L}_{fix}=-{\frac{1}{2{\alpha}}}(n^{\mu}A^a_{\mu})^2,  \ \ \
   {\alpha}{\rightarrow}0.
$$
where $A^a_{\mu}$ and $n^{\mu}$ represent the gauge boson vector field and
one light cone base vector respectively. There are many features concerning
the use of this gauge in quantum field theory \cite{bass}, but we are 
interested just in how ghots fields decoupled from theory. We are going to 
show two ways of demonstrate this fact. 
\subsection{Ghost decoupling.}
Ghost fields can be generated from the Fadeev-Popov mechanism for a
gauge vector field $A_\mu$, this means that we have to analyze the 
following path integral
\begin{equation}
        {\int}[dA_{\mu}]f(A_{\mu})e^{i{\int}d^4x{\cal L}}.
\label{b1}
\end{equation}
In fact,  $A_{\mu}$, means,  $A_{\mu}^a$,  where $a$ represents
the gauge group index. Also the functional $f(A_{\mu})$ is a gauge
invariant quantity. It is assumed  that $[dA_{\mu}]=[dA_{\mu}^{\omega}]$.
And the ansatz for $[dA_{\mu}]$ is: 
$$[dA_{\mu}]={\prod_{{\mu},a,x}}dA_{\mu}^a(x)$$
The paht integral (\ref{b1}) considers all possible configurations of
$A_{\mu}(x)$, so there is an over counting in a gauge equivalence classes.
This leads us to divide the configuration space $\{A_{\mu}(x)\}$ into
classes of equivalence $\{A_{\mu}^{\omega}(x)\}$  called gauge group orbits. 
The gauge groups orbits have all of field's configurations which results 
from applying all transformations ${\omega}$ of the gauge group ${\cal G}$
from an initial configuration $A_{\mu}(x)$ of the field.\\
The gauge transformation is
\begin{eqnarray}
        A'^a_{\mu}\ {\equiv}\ (A^{\omega})^a_{\mu}&=&
        A^a_{\mu}+f^a_{bc}A^b_{\mu}{\omega}^c+{\partial}_{\mu}{\omega}^a.
\label{bgauge}
\end{eqnarray}
Now let's allow $[d{\omega}]$ to represent an invariant measure over the 
gauge group ${\cal G}$,  i.e.  $[d{\omega}]=[d{\omega}{\omega}']$, 
where the ansatz is:
$$
        [d{\omega}]={\prod_x}d{\omega}(x).
$$
Introducing the functional ${\Delta}[A_{\mu}(x)]$ 
\begin{equation}
        1={\Delta}[A_{\mu}(x)].{\int}[d{\omega}]{\delta}{\bigl[}F[A_{\mu}^{\omega}(x)]
        {\bigr]}.
\label{b2}
\end{equation}
Where ${\delta}[f(x)]$ represents ${\prod_x}{\delta}[f(x)]$. \
Also, for $F[A_{\mu}^{\omega}]$ it is assumed:
$$
        F[A_{\mu}^{\omega}]=0.
$$
Which has exactly one solution, ${\omega}_0$, whoever is $A_{\mu}$.
This last expression is the constraint, that defines the hypersurface and
also the ``gauge''.
\begin{description}
\item[Observation:]${\Delta}[A_{\mu}(x)]$  \  is a gauge invariant since:
$$
  {\Delta}^{-1}[A_{\mu}^{\omega}]={\int}[d{\omega}']{\delta}{\bigl[}
  F[A_{\mu}^{{\omega}{\omega}'}]{\bigr]}
 ={\int}[d{\omega}{\omega}']{\delta}{\bigl[}F[A_{\mu}^{{\omega}{\omega}'}]{\bigr]},\\
  {\int}[d{\omega}"]{\delta}
  {\bigl[}F[A_{\mu}^{{\omega}"}]{\bigr]}={\Delta}^{-1}[A_{\mu}],
$$
\end{description}
inserting (\ref{b2}) into (\ref{b1}), we obtain:
$$
        {\int}[d{\omega}]{\int}[dA_{\mu}]f(A_{\mu}){\Delta}[A_{\mu}(x)]{\delta}{\bigl[}F[A_{\mu}^{\omega}(x)]
        {\bigr]}e^{iS[A_{\mu}^{\omega}]}.
$$
\begin{description}
\item[Observation:]The expression inside of ${\int}[d{\omega}]$ is also a gauge
invariant; i.e.  it does not depend on ${\omega}$.
\end{description}
\begin{equation}
        {\underbrace{{\bigl(}{\int}[d{\omega}]{\bigr)}}_{\mbox{vol.infinito}}}
        {\int}[dA_{\mu}]f(A_{\mu}){\Delta}[A_{\mu}(x)]{\delta}{\bigl[}F[A_{\mu}^{\omega}(x)]{\bigr]}
        e^{iS[A_{\mu}^{\omega}]}.
\label{b3}      
\end{equation}
{\bf Faddeev-Popov determinat.}\\
From eq. (\ref{b2}):
$$
   {\Delta}^{-1}[A_{\mu}]={\int}[dF]{\biggl(}\mbox{det}
   {\frac{{\delta}F[A_{\mu}^{\omega}]}{{\delta}{\omega}}}{\biggr)}^{-1}{\delta}F,
$$
i.e.
$$
        {\Delta}[A_{\mu}]=\mbox{det}
        {\frac{{\delta}F[A_{\mu}^{\omega}]}{{\delta}{\omega}}}
        {\bigg|}_{F[A_{\mu}^{\omega}]=0}=\mbox{det}M,
$$
${\Delta}[A_{\mu}]$ is usually called as the Faddeev-Popov determinat. If we consider
the gauge $F[A_{\mu}^{\omega}]-C(x)=0$, then our integral expression (\ref{b3}) 
shall looks like:
\begin{equation}
  {\int}[d{\omega}]{\int}[dA_{\mu}]\mbox{det}M.
  f(A_{\mu}){\delta}{\bigl[}F[A_{\mu}^{\omega}(x)]-C(x)
  {\bigr]}e^{iS[A_{\mu}^{\omega}]}.
\label{int}
\end{equation}
The expression for the functional generator Z[J=0] will be:
$$
  Z[J=0]=N{\int}[d{\omega}]{\int}[dA_{\mu}]\mbox{det}M.
  f(A_{\mu}){\delta}{\bigl[}F[A_{\mu}^{\omega}(x)]-C(x)
  {\bigr]}e^{iS[A_{\mu}^{\omega}]}.
\label{bZ}
$$
Using the matrix identity from Grassmanian numbers:
\begin{equation}
        \mbox{det}M={\int}[d{\bar{c}}][dc]{\cdot}e^{i{\bar{c}}Mc},
\label{boo}
\end{equation}
shall made possible to write $Z[J=0]$ as
\begin{eqnarray*}
        Z[J=0]&=&N{\int}[d{\omega}]{\int}[dA_{\mu}][d{\bar{c}}][dc]{\cdot}
        f(A_{\mu}){\delta}{\bigl[}F[A_{\mu}^{\omega}(x)]-C(x){\bigr]}
        e^{i(S[A_{\mu}^{\omega}]-{\bar{c}}Ac)},
\end{eqnarray*}
where the new fields $c$ and ${\bar{c}}$ are called {\it ghost fields},
and since its statistical nature, they just appear as internal lines in 
Feynman graphs.\\
With the preliminaries given above now we are ready to show the ghosts 
decoupling. We will have two ways of demonstrate how the ghosts 
decuple in an axial type gauge:
\subsubsection{\bf Ghosts decoupling: way A.}
Starting from the axial type gauge definition
\begin{eqnarray*}
        n^{\mu}A^a_{\mu}&=&0,\\
        n^{\mu}n_{\mu}&=&\mbox{const},
\end{eqnarray*}
the fixing gauge term is $F^a=n^{\mu}A^a_{\mu}.$
Then, making use gauge transformation (\ref{bgauge}) we have
$$
        {\delta}F^a\ {\equiv}\ n^\mu f^{abc}A^b_{\mu}{\omega}^c+
        n^\mu{\partial}_{\mu}{\omega}^a.
$$
This means that
$$
        {\frac{{\delta}F^a}{{\delta}{\omega}_b}}
        ={\delta}^{ab}n^{\mu}{\partial}_{\mu}.
$$
Observe that the last expression represents the matrix $M$ and for this case 
it does not involve the gauge field $A^a_{\mu}$, 
this means that also $\mbox{det}M$ does not
have $A^a_{\mu}$ and this makes possible to put $\mbox{det}M$ out from the 
path integral of $[dA_{\mu}]$ in (\ref{int}). Then the expression for $Z[J=0]$
should be written as
$$
        Z[J=0]=N{\int}[d{\omega}]detM.{\int}[dA_{\mu}]
        f(A_{\mu}){\delta}{\bigl[}F[A_{\mu}^{\omega}(x)]-C(x){\bigr]}
        e^{iS[A_{\mu}^{\omega}]}.
$$
The term ${\int}[d{\omega}]\mbox{det}M$ could be absorved into the
constant, or if you prefer, we shall make a redefinition, nevertheless, 
arriving to
$$
        Z[J=0]=N{\int}[dA_{\mu}]f(A_{\mu}){\delta}{\bigl[}
        F[A_{\mu}^{\omega}(x)]-C(x){\bigr]}e^{iS[A_{\mu}^{\omega}]}.
$$
As we can see, this last expression is ghosts fields free, that is 
the point we wanted to demonstrate. 
\subsubsection{\bf Ghosts decoupling: way B.}
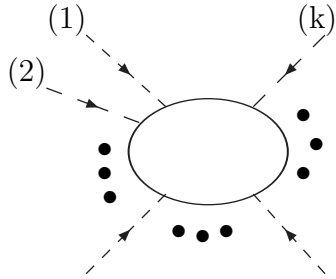
\begin{figure}[ht] 
\vskip 0.5in
\centerline{
\begin{picture}(120,90)(-60,-30)
\Oval(16,16)(20,30)(0)
\DashArrowLine(-30,60)(0,33)5
\put(-30,60){\makebox(0,0)[br]{(1)}}
\DashArrowLine(-45,40)(-10,27)5
\put(-45,40){\makebox(0,0)[br]{(2)}}
\put(-20,14){\makebox(0,0)[br]{${\bullet}$}}
\put(-20,5){\makebox(0,0)[br]{${\bullet}$}}
\put(-18,-4){\makebox(0,0)[br]{${\bullet}$}}
\DashArrowLine(-30,-30)(0,0)5
\put(8,-17){\makebox(0,0)[br]{${\bullet}$}}
\put(17,-19){\makebox(0,0)[br]{${\bullet}$}}
\put(26,-17){\makebox(0,0)[br]{${\bullet}$}}
\DashArrowLine(60,-30)(33,0)5
\put(55,27){\makebox(0,0)[br]{${\bullet}$}}
\put(60,16){\makebox(0,0)[br]{${\bullet}$}}
\put(55,5){\makebox(0,0)[br]{${\bullet}$}}
\DashArrowLine(60,60)(33,33)5
\put(65,60){\makebox(0,0)[br]{(k)}}
\end{picture}
}
\caption{\label{nghost} $k-$points ghost function.}
\end{figure}

Starting from the definition of $\mbox{det}M$  and using the property of
the product of matrix determinats:
\begin{eqnarray*}
        \mbox{det}M&=&\mbox{det}(n^{\mu}D_{\mu}),\\
        &=&\mbox{det}(n.{\partial}){\cdot}\mbox{det}(1-gGf^aA^{a{\mu}}n_{\mu}),
\end{eqnarray*}
where in this last expression we are using the adjoint representation. We also
have that
\begin{eqnarray*}
        D_{\mu}&=&{\partial}_{\mu}-gf^an_{\mu}{\partial}^{\nu}A^a_{\nu},\\
        ({\partial}.n)G(x-y)&=&{\delta}^D(x-y),
\end{eqnarray*}
so, working with the second determinant, using another well-known property
of matrix determinats and then expanding the logarithmics in a sum over
$k\ (k=1{\cdots}{\infty})$:
\begin{eqnarray*}
        \mbox{det}(1-gGf^aA^a_{\mu}n_{\mu})&=&
        e^{Tr[{\ln{(1-gGf^aA^a_{\mu}n_{\mu})}}]},\\
        &=&e^{-Tr{\big[}{\sum}{\frac{g^k}{k}}
        G(x_0-x_1)f^{a_1}A^{a_1}_{{\mu}_1}n_{{\mu}_1}{\cdots}
        G(x_{k-1}-x_k)f^{a_k}A^{a_l}_{{\mu}_k}n_{{\mu}_k}{\big]}}.
\end{eqnarray*}
Here we observe that we obtain a serie of graphs, that are functions of 
$k-$points, in other words, each term of the sum is a Feynman graph that
can be represented in the figure (\ref{nghost}).\\
%
%
This implies that each process involves Feynman integrals, which 
can be computed (here we are using the Feynman parametrization) in the context
of dimensional regularization. In this way, the integral that is associated 
to the figure (\ref{nghost}) will have the form:
\begin{eqnarray*}
        I^{a_1{\cdots}a_k}_{{\mu}_1{\cdots}{\mu}_k}
        &=&g^k{\prod^k_{i=1}}n_{{\mu}_i}Tr{\Big[}{\prod^k_{i=1}}f^{a_i}{\Big]}(k-1)!
        \int\limits^1_0dx_1{\cdot}\int\limits^{x_{k-1}}_0dx_{k-2}
        {\int}{\frac{d^Dq}{(n.q+{\cdots}-n.p_2x_{k-1})^k}}.
\end{eqnarray*}
Making a change of variables,
$$
        I^{a_1{\cdots}a_k}_{{\mu}_1{\cdots}{\mu}_k}
        =g^k{\prod^k_{i=1}}n_{{\mu}_i}Tr{\Big[}{\prod^k_{i=1}}f^{a_i}{\Big]}I.
$$
Where $I$ (using Lorentz invariance)
\begin{eqnarray*}
        I&=&{\int}{\frac{d^Dq}{(n.q)^k}},\\
        &=&c(n,k){\frac{1}{n^k}}{\int}{\frac{d^Dq}{q^k}}.
\end{eqnarray*}
This last integral is zero in the context of dimensional regularization.
With this in mind it is easy to see that $\mbox{det}(1-gGf^aA^a_{\mu}n_{\mu})=1$,  
this means that $\mbox{det}M$ again does not depend on ghost fields, so again
we are able to say that the ghost fields decoupled from the theory. 
\section{\large Complete two light cone base vectors.}
The four Minkowski space-time can be generated using a base of just two vectors:
the light cone vector.
Defining the dual base light-like four-vectors:
\begin{eqnarray}
  \nonumber
  n_{\mu} &=& {\frac{1}{\sqrt{2}}}(1,0,0,1),\\
  m_{\mu} &=& {\frac{1}{\sqrt{2}}}(1,0,0,-1),
\end{eqnarray}
we observe that with the help of this base, the $x^{\pm}$ coordinates
can be expressed as
\begin{eqnarray*}
        x^+ &=& x^{\mu}n_{\mu},\\
        x^- &=& x^{\mu}m_{\mu}.
\end{eqnarray*}
Using these two light-cone vectors, it is possible to built up a two
degree light cone gauge theory as presented in \cite{lpg}. In this 
case, the suitable gauge fixing Lagragian shall be
\begin{equation}
  \label{lpg}
  {\cal L}_{fix}=-\frac{1}{2\alpha}(n.A)(m.A).
\end{equation}
this term has the disadvantage of generate a more complicate structure
of boson propagator in the theory, but allows us to discern some 
interesting properties such us the possibilities of defining a light like
planar gauge and  have a prescriptionless theory \cite{lcgML} for the so called
spurious poles. In this work we will show anhoter new feature. Let's 
first see the ghost behavior for this case.
\subsection{Ghost non decoupling.}
From eq. (\ref{bgauge}) we have that the variation of the gauge 
field is:
$$
  \delta A^a_\mu=f^a_{bc}A^b_{\mu}{\omega}^c+{\partial}_{\mu}{\omega}^a,
$$
and the Fadeev-Popov matrix is:
$$
  M={\frac{{\delta}F[A_{\mu}^{\omega}]}{{\delta}{\omega}}}
    {\bigg|}_{F[A_{\mu}^{\omega}]=0} 
$$
Since the gauge fixing Lagrangian is shown in eq. (\ref{lpg}), we have that 
in this case the gauge fixing term is
\begin{equation}
  \label{gf}
  F^a=[(n.A^a)(m.A^a)]^{1/2},
\end{equation}
then is variation should take the form
\begin{eqnarray*}
  \delta F^a=&\frac{1}{2[(n.A^a)(m.A^a)]^{1/2}}&
             \{(n^\mu f^a_{bc}A^b_\mu\omega^c+n^\mu\partial_\mu\omega^a)(m.A^a)\\
             &&+(n.A^a)(m^\mu f^a_{bc}A^b_\mu\omega^c+m^\mu\partial_\mu\omega^a)\},
\end{eqnarray*}
that allows us to find
\begin{equation}
  \label{final}
  \frac{\delta F^a}{\delta\omega_b}=\frac{\delta^{ab}}{2[(n.A^a)(m.A^a)]^{1/2}}
                  \{(n^\mu\partial_\mu)(m.A)+(n.A)(m^\mu\partial_\mu)\}.
\end{equation}
Actually eq. (\ref{final}) represents the Fadeev-Popov matrix. It says that
ghost {\bf does not decuple} here, so they have a dynamical role in
this axial type gauge. 
\section{\large Conclusions.}
The fact that ghosts does not decuple from theory has a very close 
relation with the BRST operator, which
now can be defined in the same manner as in eq. (\ref{BRST}) as
$$
  Q=c^aG_a-\frac{i}{2}f_{ab}^cc^ac^bb_c,
$$
where the $G_a$ are generators of the Lie algebra associated to
Yang-Mills theory on the light-cone gauge in four dimensions.
%
The existence of light-like planar gauge is not an oldest one. In fact it 
was belived that there was not a consistent form to define it, 
until studies of some kind of axial type gauges with two degree of 
freedom \cite{us}: we are talking about the use of both light-cone base vectors, 
which opened a small track in non covariant gauges studies. 
Besides this gauge, as we alredy have seen, takes care of the 
proper role of ghosts and its importance, in particular on the 
BRST-noncovariant gauges question.
We are not boasted about this kind of gauge is better that the other one,
we just want to point out this new feature (consistent BRST operator 
for the light cone gauge) when working with the full base of 
light cone vectors. 
\newpage
{\bf Acknowledgments:} We are very grateful to both the Departmento de 
Matem\'atica and the Departamento de F\'{\i}sica for their hospitality 
at Universidade Federal de S\~ao Carlos where this work was prepared.
One big special word of grateful to professor H. Abdalla Neto who suggest 
me the possibility of study this topic. This work is dedicated to
my sister Dana and my brother Daniel on their birthday.\\

\end{document}